\documentclass[12pt]{iopart}
\usepackage{iopams}
\usepackage{graphicx}
\begin{document}

\title[Thermal conductivity and stability of commercial MgB$_2$ conductors]{Thermal conductivity and stability of commercial MgB$_2$ conductors}

\author{Marco Bonura and Carmine Senatore}

\address{Department of Applied Physics (GAP) and Department of Quantum Matter Physics (DQMP), University of
Geneva, quai Ernest Ansermet 24, CH-1211 Geneva, Switzerland}
\ead{marco.bonura@unige.ch}
\begin{abstract}

This paper presents a study of the thermal transport properties of MgB$_2$ tapes differing in architecture, stabilization and constituent materials. The temperature and field dependence of thermal conductivity, $\kappa(T,B)$, was investigated both along the conductor and in the direction perpendicular to the tape. These data provide fundamental input parameters to describe the 3D heat diffusion process in a winding. Thermal transport properties - even in field - are typically deduced using semi-empirical formulas based on the residual resistivity ratio of the stabilizer measured in absence of magnetic field. The accuracy of these procedures was evaluated comparing the calculated $\kappa$ values with the measured ones. Based on the experimental thermal conduction properties $\kappa(T,B)$ and critical current surface $J_C(T,B)$ we determined the dependence of minimum quench energy and normal zone propagation velocity on the operating parameters of the conductor. The correlation between thermal properties and tape layout allowed us to provide information on how to optimize the thermal stability of MgB$_2$ conductors.

\end{abstract}

\maketitle

\section{Introduction}
Since the discovery of superconductivity in MgB$_2$, it has been clear that magnesium diboride can represent a valid solution for many practical applications of superconductivity. The main points of strength of this compound are its critical temperature of nearly 40~K, which may allow operating in cryogen-free environments, and its cost. Indeed, estimations based on manufacturing and raw material costs indicate that the price per ampere-meter can be lower than that of NbTi, which is presently the least expensive technical superconductor \cite{Tomsic}. MRI magnets are considered the largest market for MgB$_2$ \cite{Ling,Razeti}. Nevertheless, other applications are currently under investigation, as for example high current cables (see e.g. the link project at CERN \cite{Sugano}), fault current limiters, transformers, motors and generators \cite{Tomsic,Lin,Nakamura}.

The thermal stability of a superconducting device is an important issue, whatever is the application. In case of thermal disturbances, the temperature of the conductor can locally raise up to values higher than $T_{CS}$, the temperature at which a sharing of the transport current between the superconductor and the metals present in the composite starts. In this case, heat is generated by Joule effect and a quench of the whole system can occur if the perturbation energy is higher than the \textit{minimum quench energy} \cite{Wilson,Iwasa}. The \textit{normal zone propagation velocity}, i.e. the velocity of the normal/superconducting boundary during a quench is another important parameter for the protection of the superconducting device. High velocities are desired since this limits the final temperature of the hot spot and makes the detection of the perturbation - and consequently the activation of the protection system - faster \cite{Wilson,Iwasa}.

The thermal conductivity, $\kappa$, is a key parameter for modeling the response of the superconductor to a thermal perturbation. Both the longitudinal, $\kappa_L$, and the transverse, $\kappa_T$, components of $\kappa$ are necessary to describe the 3D heat diffusion process in a winding. $\kappa_L$ and $\kappa_T$ result from the conductor architecture and materials. It is expected that  in thermally stabilized superconductors the predominant contribution to $\kappa_L$ originates from the stabilizer, typically copper \cite{Wilson,Bonura1}. Thermal conduction in metals at cryogenic temperatures is mainly determined by electron-defect scattering processes and thus purer metals better conduct heat \cite{Hust}. The magnetic field lowers the electron mobility and, consequently, the thermal conductivity \cite{Bonura1,Hust}. This makes important the study of the field dependence of the thermal properties especially in view of the use of the wires in high field magnets \cite{Mine}.

This manuscript presents a study of the thermal transport in MgB$_2$ conductors differing in architecture, stabilization and constituent materials. The accuracy of the analytical procedures generally used to estimate in-field $\kappa_L$ from an electrical characterization of the stabilizer is evaluated. The thermal conduction has also been measured in the direction perpendicular to the tape. From the experimental investigation of the thermal conduction properties, we determine the dependence on the operating conditions of the minimum quench energy and normal zone propagation velocity, fundamental parameters in the definition of the stability margins of SC-based devices. The analysis of the correlations between thermal properties and conductor layout allows us to discuss solutions to make the conductor design more suitable for specific applications.

\section{Samples}\label{Samples}
\begin{figure}[t]
\centering \includegraphics[width=8 cm]{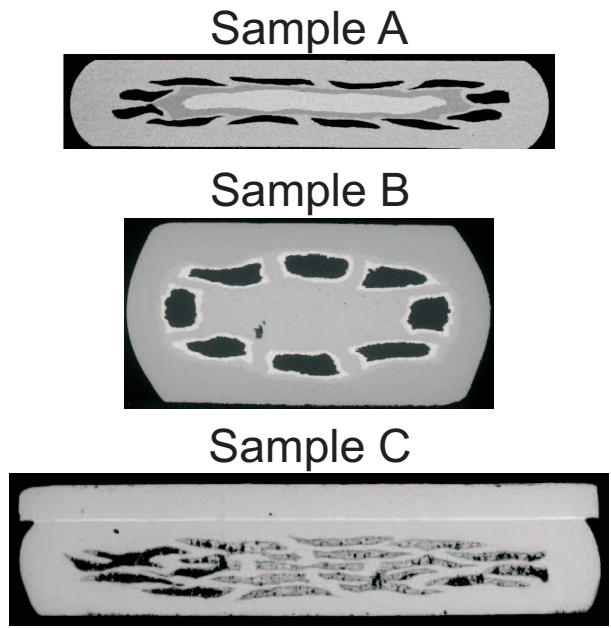} \caption{\label{Fig1}
Pictures of the cross sections of the MgB$_2$ conductors. Sample A is composed of 12 filaments of MgB$_2$ embedded in a Ni matrix. It is stabilized by a Cu core coated with pure Fe. Sample B is a 8-filament conductor with an inner Cu core shielded with CuNi/Nb. The matrix is in stainless steel. Sample C has 19 MgB$_2$ filaments embedded in a Ni matrix and is stabilized via an external Cu strip.}
\end{figure}

Three MgB$_2$ tapes with different layout produced by Columbus Superconductors were investigated. The conductor cross sections are shown in Figure~\ref{Fig1}, while the maximum width and thickness of the tapes are listed in Table~\ref{TabSamples1} along with other structural parameters as the number of filaments and the type of stabilization. Sample~A is composed of 12 filaments of MgB$_2$ and has a Ni matrix. It is stabilized by a Cu core coated with pure Fe in order to prevent chemical poisoning during the heat treatment. Sample~B is a 8-filament conductor with an inner Cu core. The stabilizer is shielded with CuNi/Nb. The matrix consists of stainless steel (SS). Ni is used as matrix material also in the case of Sample C. This conductor presents 19 partly bridged MgB$_2$ filaments. A Cu strip of thickness $\approx 0.2$~mm is soldered on one side of the tape for thermal stabilization. No barrier is needed for chemical protection of the stabilizer with this architecture. The volume fraction occupied by each constituent material is listed in Table~\ref{TabSamples2}.

The critical temperature of the conductors, defined at half of the resistive transition, is $T_C \approx 38.5$~K, $\approx 37.0$~K, $\approx 35.5$~K in Samples A, B and C, respectively. The width of the transition, defined as the difference between the temperatures at the 90\% and 10\% of the superconducting transition is $\Delta T_C\approx 0.3$~K, $\approx 0.4$~K and $\approx 0.4$~K for Samples A, B and C, respectively. Critical current values at $T=20$~K and $B=1$~T are $I_C\approx 345$~A for Sample~A, $\approx 240$~A for Sample~B, and $\approx 300$~A for Sample~C. These values, normalized to the conductor cross sections, lead to comparable engineering critical current densities for Samples~A and C, which have $J_{eng}\approx 150$~A/mm$^2$. The engineering critical current density is lower for Sample~B, being $J_{eng}\approx 100$~A/mm$^2$.

\begin{table}[t]
\caption{\label{TabSamples1} Geometrical characteristics of the MgB$_2$ conductors}
\footnotesize\rm
\begin{tabular*}{\textwidth}{@{}l*{15}{@{\extracolsep{0pt plus12pt}}l}}
\br
 & Sample A & Sample B & Sample C \\
\mr
Width (mm) & 3.65 & 2.25 & 3.00\\
Thickness (mm)  & 0.65 & 1.10 & 0.65\\
N. of filaments      & 12 & 8 & 19\\
Cu stabilization      & Internal & Internal & External\\
\br
\end{tabular*}
\end{table}

\begin{table}[t]
\caption{\label{TabSamples2} Composition of the MgB$_2$ conductors}
\footnotesize\rm
\begin{tabular*}{\textwidth}{@{}l*{15}{@{\extracolsep{0pt plus12pt}}l}}
\br
 & Sample A & Sample B & Sample C \\
\mr
MgB$_2$ & 12\% & 13\% & 16\%\\
Cu      & 15\% & 14\% & 27\%\\
Ni      & 63\% & n.a. & 57\%\\
Fe      & 10\% & n.a. & n.a.\\
CuNi/Nb & n.a. & 26\% & n.a.\\
Stainless Steel & n.a. & 47\% & n.a.\\
\br
\end{tabular*}
\end{table}
\section{Experimental technique}\label{Experimental technique}
The temperature and field dependence of the thermal conductivity were investigated by means of a setup specifically designed for measurements on technical superconductors. The system allows measuring both the longitudinal and transverse thermal conductivities in magnetic fields up to 21~T \cite{Bonura1,Bonura2,Bonura3}.

\begin{figure}[t]
\centering \includegraphics[width=8 cm]{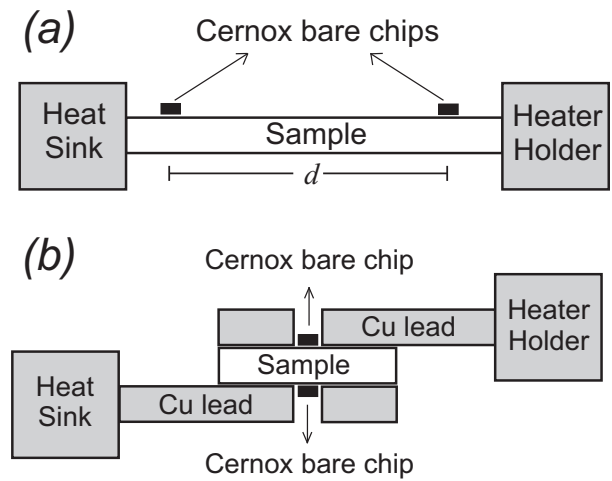} \caption{\label{Fig2}
Sketch of the sample holder for longitudinal \textit{(a)} and transverse \textit{(b)} thermal conductivity measurements.}
\end{figure}

During the measurement, heat $Q$ supplied to the sample creates a temperature gradient $\Delta T$ along it.  $\kappa_L$ and $\kappa_T$ are defined as:
\begin{eqnarray}\label{k}
    \kappa_{L}=\frac{Q}{\Delta T}\cdot \frac{d}{w\cdot t} ,\ \ \
    \kappa_{T}=\frac{Q}{\Delta T}\cdot \frac{t}{w\cdot l} \ ,
\end{eqnarray}
where $l, w, t$ are the length, width and thickness of the sample, respectively. As shown in Figure~\ref{Fig2}, $d$ is the distance between the temperature taps in the longitudinal case.

Different sample mounting procedures are required for measuring $\kappa_L$ and $\kappa_T$. In Figure~\ref{Fig2}, we report a schematic drawing of the sample holder for longitudinal (\textit{a}) and transverse (\textit{b}) measurements. In case (\textit{a}), heat power is supplied to one end of the tape, ensuring a good thermal contact with the heater over the entire conductor cross section. Thus, all materials present in the tape contribute to the thermal transport. The temperature gradient along the sample is measured by two Cernox bare chips directly glued on the sample surface. In case (\textit{b}), the sample is sandwiched between two copper leads glued on the sample faces by GE varnish. This ensures good thermal contact over the entire tape surface. The leads are in high-$RRR$ Cu and the temperature gradient along them is negligible with respect to $\Delta T$. The Cernox bare chips, used as thermometers, are hosted in holes present in the Cu leads in correspondence of the center of the sample and are glued directly on the tape faces. Two calibrated Cernox thermometers are also present in the setup, one on the heat sink and one on the heater holder. During the measurement, the heat-sink temperature is kept constant. When current is supplied to the heater, energy is generated by Joule effect and a temperature gradient develops along the sample or between the two sample faces in case of longitudinal and transverse measurements, respectively. $\Delta T$ is measured once the steady-state heat flow has been reached. Typically $\Delta T$ spans from $\sim1$~mK to $\sim100$~mK as a function of the sample properties and of the type of measurement (longitudinal or transverse).
The bare chips are calibrated during the measurement, when the sample is at the thermal equilibrium with the bath.

The data uncertainty depends on the type of measurement and is mostly due to inaccuracy in assessing the distance between the thermometers. It is $\approx 5 \%$ for longitudinal measurements and $\approx 30 \%$ in the transverse case. The use of Cernox bare chips as thermometers guaranties very high sensitivities, especially at low temperatures ($T\lesssim10$~K) when high values for the derivative of the resistance vs temperature curve ($dR/dT$) permit a detection of temperature gradients of $\sim 1$~mK. The value of $dR/dT$ is not constant, and this reduces the signal-to-noise ratio on increasing the temperature.

\section{Experimental Results}\label{Results}
The thermal conductivity has been investigated in the range of temperatures from 5~K to 40~K at $B=0$ and $B=7$~T, this field value being an upper limit for MgB$_2$ conductors in typical applications \cite{Budko,Collings}. $B$ has been applied in the plane of the tape. $\kappa_L$ measurements have been performed with the magnetic field both parallel and perpendicular to the thermal current direction, the latter configuration being more relevant for coils. In case of transverse thermal conductivity measurements, the magnetic field is always perpendicular to the thermal current with our setup. We have measured an extra datapoint of $\kappa_L$ at $T\approx 78$~K and $B=0$ inserting the probe in a container filled with liquid nitrogen.

\begin{figure}[t]
\centering \includegraphics[width=8 cm]{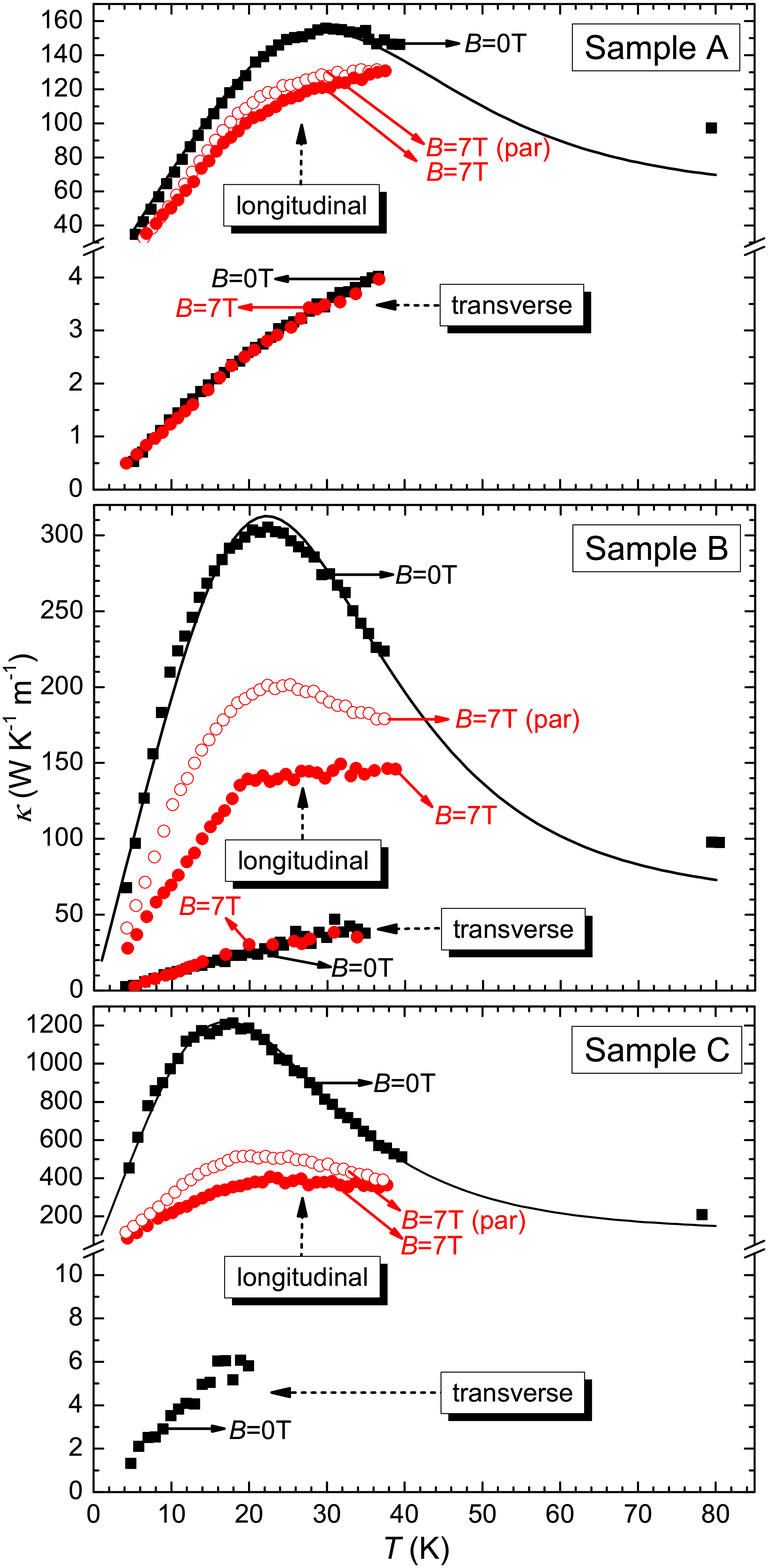} \caption{\label{Fig3}
Temperature dependence of the longitudinal and transverse thermal conductivities of MgB$_2$ conductors. Longitudinal $\kappa$ data acquired with the magnetic field applied parallelly to the thermal current direction are shown as open dots.}
\end{figure}

In Figure~\ref{Fig3} it is shown the experimental temperature dependence of the longitudinal and transverse thermal conductivities. The conductor layout has a major influence on the thermal response. The best thermal conduction properties have been observed in Sample~C, which presents an external stabilization. $\kappa_L(T)$ curves at $B=0$ present a maximum whose value and position depend on the sample.  The maximum occurs at $T\approx 30$~K, $\approx 22$~K and $\approx 17$~K in Sample~A, B and C, respectively. $\kappa_L(B=0)$ at the maximum spans over one order of magnitude among the samples. The smallest value, $\kappa_L\approx150$~WK$^{-1}$m$^{-1}$, has been measured in Sample A; the largest, $\kappa_L\approx1200$~WK$^{-1}$m$^{-1}$, in Sample~C. A correlation exists between peak position and $\kappa_L$ magnitude, the best thermal conduction properties being realized in the sample with the lowest $T$ of the peak. $\kappa_L$ decreases on applying the magnetic field. Field-induced effects strongly depend on the conductor properties and are particularly intense for Sample~C. As a consequence, the differences between $\kappa_L$ values measured at $B=0$ in the three samples are considerably reduced after the application of the magnetic field. At temperatures higher than $\sim40$~K, electron-phonon scattering processes start becoming predominant in determining $\kappa$. At lower temperatures, the largest contribution comes from electron-defect scattering events \cite{Hust}. It follows that increasing the purity of the metal is effective in promoting the thermal conduction only at low temperatures. The effect of the magnetic field is in some way analogous to that of disorder in the system: both reduce the electron mean free path and consequently $\kappa$. The largest variation of $\kappa$ is found when the magnetic field is applied perpendicularly to the thermal current direction. Indeed, in this configuration the reduction of the mean free path due to the action of the Lorentz force on the electrons is maximized. At temperatures higher than $\sim 40$~K, neither the purity nor the magnetic field should affect noticeably the thermal transport. This is the reason why zero-field and in-field $\kappa(T)$ curves approach each other on increasing the temperature, as shown in Figure~\ref{Fig3}.

Transverse thermal conductivity values are much smaller than the longitudinal ones. This is a consequence of the design of the conductors that promotes heat diffusion along the tape. $\kappa_T$ increases monotonically with the temperature in the whole range of temperatures investigated. Values measured in Sample~A and Sample~C are lower than 10~WK$^{-1}$m$^{-1}$. In Sample~B, we have measured $\kappa_T$ up to about 45~WK$^{-1}$m$^{-1}$. In Samples A and B we have also investigated the effect of the magnetic field on the transverse thermal conduction. We have found that, contrary to what observed for the longitudinal heat transfer, the effect of the field on the thermal conduction across the wire cannot be resolved. This indicates that the Cu contribution to the overall transverse thermal resistance is negligible, as already observed in REBCO coated conductors \cite{Bonura3}. We have not measured in-field $\kappa_T$ in Sample~C. Geometrical and thermal properties of this sample make the signal-to-noise ratio low, especially for $T\gtrsim 15$~K when the sensitivity of the Cernox bare chips becomes smaller and affects the accuracy of the measurement. This is also the reason why $\kappa_T$ measurement at $B=0$ was limited for this sample to a maximum temperature of about 20~K.

%


\section{Discussion} \label{Discussion}
\subsection{Longitudinal Thermal Conductivity at $B=0$} \label{Discussion-k_l(0)}
Longitudinal thermal transport in composite conductors can be examined with a formalism analogous to the case of electrical resistances connected in parallel. The overall thermal conductivity is the weighted sum of the thermal conductivity of each component, $\kappa_i$, with weights $s_i \equiv S_i / S_{tot}$, where $S_i$ and $S_{tot}$ are the surface of the cross section occupied by the $i^{th}$ component and the total cross-section area of the conductor, respectively: $\kappa_{L}=\sum \kappa_i s_i$.

In metals, thermal conduction occurs via the electronic and lattice channels, the former being the most relevant one \cite{Hust}. The electronic contribution is in turn determined by electron-defect and electron-phonon scattering processes. Thus, the temperature dependence of the thermal conductivity of metals is modeled as:

\begin{eqnarray}\label{k_Cu}
    \kappa_{Cu}=(W_0+W_i+W_{i0})^{-1} ,\\
    W_0= \frac{\beta}{T} ,\nonumber \\
    W_i=\frac{P_1 T^{P2}}{1+P_1 P_3 T^{(P_2+P_4)}e^{-(P_5/T)^{P_6}}} ,\nonumber \\
    W_{i0}=P_7\frac{W_i W_0}{W_i+W_0} , \nonumber
\end{eqnarray}

where $W_0$ and $W_i$ represent the electron-defect and electron-phonon contribution, respectively. $W_{i0}$ is an interaction term between $W_0$ and $W_i$ \cite{Hust}. $\beta$ is a function of the residual resistivity ratio, $RRR\equiv \rho$(273~K)/$\rho_{res}$, where $\rho_{res}$ is the residual electrical resistivity measured at low temperature. The parameters $\beta$ and $P_i$ differ from one metal to another. They have been determined by least squares fit of the experimental data. For Cu, the values are: $\beta\approx 0.634 / RRR$, $P_1 = 1.754 \times10^{-8}$, $P_2 = 2.763$, $P_3 = 1102$, $P_4 = -0.165$, $P_5 = 70$, $P_6 = 1.756$, $P_7 =0.235 \cdot RRR ^{0.1661}$ in SI units \cite{Hust}. At temperatures lower than $\approx 50$~K the electron-defect term, $W_0$, is predominant and $\kappa$ is mainly determined by the $RRR$ of the metal. In particular, samples with higher $RRR$ are more efficient in conducting heat.

We expect that the main contribution to $\kappa_L$ in the MgB$_2$ conductors investigated in this work comes from Cu. This is a consequence of thermal conduction properties of the constituent materials \cite{Hust,Bauer,Putti,Bauer2,Kemp} and of the fact that only the stabilizer has been optimized for heat conduction, whilst the other metals present in the tape have $RRR$ lower than $\approx 10$ \cite{Tropeano}. In particular, we do not expect to see any noticeable effect on the longitudinal thermal transport related to the MgB$_2$ phase in the range 5~K$ \div $40~K. Thermal conduction properties of MgB$_2$ have been reported in the literature. $\kappa(T)$ curves show a peak between 60~K and 110~K. The exact position of the maximum as well as the $\kappa$ values depend on the sample purity and density \cite{Putti,Bauer2}. $\kappa$ of the order of $\sim 10$ Wm$^{-1}$K$^{-1}$ is expected in the case of MgB$_2$ filaments in a conductor \cite{Bauer2}. $\kappa\sim 10$ Wm$^{-1}$K$^{-1}$ is approximatively two order of magnitudes smaller than values measured in Cu with $RRR=30$ in the same $T$ range \cite{Hust}. As a consequence, we have analyzed the experimental results obtained at $B=0$ in the approximation $\kappa_L\approx \kappa_{Cu}\cdot s_{Cu}$ using Equation~\ref{k_Cu}, $s_{Cu}$ values reported in Table~\ref{TabSamples2}, and keeping the $RRR$ of Cu as the only free fitting parameter. The best fit curves are shown in Figure~\ref{Fig3} as continuous lines. A very good agreement between experimental and calculated curves has been obtained in the whole range of temperatures investigated. Data measured at liquid-nitrogen temperature are up to about 30$\%$ higher than calculated values. This could be due to non-Cu contributions to the overall thermal conductivity observable at high temperatures, when $\kappa_{Cu}$ becomes smaller.

$RRR$s resulted from the best-fit procedure are listed in Table~\ref{Tab3}. A measurement of the temperature dependence of the electrical resistivity performed on the superconducting tape does not allow one to extract the $RRR$ of the stabilizer, the superconducting transition preventing the measurement of the value $\rho_{res}$ of Cu. In order to circumvent this problem and check the consistency of the best-fit parameter, we have measured the electrical resistivity with a standard four-probe technique on Cu specimens extracted from the MgB$_2$ tapes. For Sample~A and Sample~B we have extracted the stabilizer from the interior of the conductor removing the other layers by a combination of chemical etching and mechanical abrasion. For Sample~C, we have unsoldered the Cu strip using a heating plate at 250$^{\circ}$C. The process lasts about 1~min, thus limiting possible annealing effects. The procedure has been repeated on several Cu samples for each MgB$_2$ conductor, finding $RRR$ variations within the $15\%$. The average values of $RRR$ obtained by electrical measurements are also shown in Table~\ref{Tab3}.  A comparison of measured and best-fit $RRR$ indicates differences $\lesssim 20\%$, and this is within the accuracy of Equation~\ref{k_Cu} \cite{Hust}. This result confirms the validity of the approximation $\kappa_L\approx\kappa_{Cu}s_{Cu}$ for the investigated MgB$_2$ conductors.

\begin{table}[t]
\caption{\label{Tab3}Electrical and thermal-stability parameters of the investigated MgB$_2$ conductors. $MQE$ and $NZPV$ have been evaluated at $T_{Op}=20$~K, $I_{Op}=100$~A, $B=1$~T. Critical current, $I_C$, measured at $T=20$~K, $B=1$~T.}
\footnotesize\rm
\begin{tabular*}{\textwidth}{@{}l*{15}{@{\extracolsep{0pt plus12pt}}l}}
\br
 & Sample A & Sample B & Sample C \\
\mr
$RRR_{Cu}$ from $\kappa(T)$ & 30 & 90 & 238\\
$RRR_{Cu}$ from $\rho(T)$ & 37 & 86 & 255\\
$MQE$ [mJ] & 6 & 21 & 34\\
$NZPV_L$ [cm/s] & 10 & 6 & 12\\
$I_C$ [A] & 345 & 240 & 300\\
\br
\end{tabular*}
\end{table}

\subsection{Longitudinal Thermal Conductivity at $B\neq0$} \label{Discussion-k_l(B)}
The electronic contribution to thermal conductivity in metals is proportional to the electron mean free path, $\ell$ \cite{Kittel}. The application of a magnetic field reduces $\ell$ and, consequently, the thermal conductivity. Not only $\kappa$ but also the electrical resistivity, $\rho$, is related to $\ell$. The Wiedemann-Franz law establishes that their product is proportional to the temperature, the constant being the Lorenz number, $L$: $\kappa\rho=LT$. $L$ is expected to be field independent if the electron scattering events are elastic \cite{Arenz,White}. In this case, in-field $\kappa$ can be deduced from the magnetoresistance of the metal \cite{Bonura1}:

\begin{equation}\label{MR}
    \kappa(T,B)= \frac{\rho(T,0)}{\rho(T,B)} \kappa(T,0) \, .
\end{equation}

The magnetoresistance of Cu has been widely reported in the literature for fields applied perpendicularly to the electrical current. Simon \textit{et al.} \cite{Simon} have collected in-field data from different references and summarized them in a unique Kohler plot where the fractional increase in resistance $\Delta\rho$($T,B$)$\equiv[\rho(T,B)-\rho(T,0)]/\rho(T,0)$ is plotted as a function of the product $B\times [\rho(273$~K,$B=0$)/$\rho(T$,$B=0$)]. The authors have derived an average Kohler curve to evaluate the in-field electrical resistivity starting from $\rho$ values measured at $B=0$.

\begin{figure}[t]
\centering \includegraphics[width=8 cm]{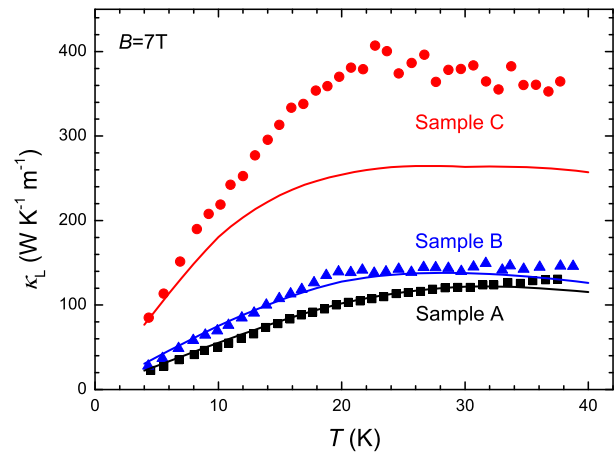} \caption{\label{Fig4}
Temperature dependence of the longitudinal thermal conductivity at $B=7$~T with the field applied perpendicularly to the thermal current direction. Points are the experimental data, lines are the theoretical curves calculated as described in the text.}
\end{figure}

Due to difficulties associated with in-field thermal conductivity measurements, $\kappa$ at $B\neq0$ is usually evaluated by Equation~\ref{MR} starting from the $RRR$ determined at $B=0$. In fact, $\kappa(T,0)$ can be calculated from the $RRR$ value by means of Equation~\ref{k_Cu}. Furthermore, the $RRR$ allows calculating the theoretical $\rho(T,B=0)$ curve of Cu and, consequently, the ratio $\rho(273$~K,$B=0$)/$\rho(T$,$B=0$) necessary to calculate the magnetoresistance from the Kohler plot \cite{Hust}. In order to evaluate the accuracy of this procedure, in Figure~\ref{Fig4} we plot the $\kappa_L$ data measured  at $B=7$~T in the perpendicular configuration, along with the theoretical lines calculated using the $RRR$ values deduced by the electrical resistivity measurements performed on the stabilizer (data shown in Table~\ref{Tab3}). The comparison between experimental and calculated results shows a very good agreement in the case of Sample~A and Sample~B, whilst calculated values for Sample~C are up to $30\%$ smaller than measured data, the largest variation having been observed at $T\approx25$~K. However, this discrepancy is within the accuracy associated with the use of Equations~\ref{k_Cu} and \ref{MR}. Indeed, from the Kohler plot one finds that $\rho(25$~K,0~T)/$\rho(25$~K,7~T)$= 0.27 \pm 0.1$ for $RRR=255$, which is the value measured for Sample~C. This corresponds to an uncertainty in assessing the magnetoresistance of about $\pm 30\%$, that has to be added to the one associated with Equation~\ref{k_Cu}. Therefore, the total uncertainty is about $\pm 50\%$.

It is noteworthy that an evident discrepancy between calculated and experimental curves has been obtained in the sample with the highest $RRR$ of Cu. This is related to the fact that the field-induced effects on $\kappa$ are more pronounced in pure metals. We would also like to underline that we have achieved a very good agreement between theoretical and experimental curves in the case of Sample~A and Sample~B because of the precise measurement of the $RRR$ of the stabilizer. The $RRR$ of raw materials is often used for evaluating the thermal properties of the composite. However, this procedure excludes all modifications resulting from the fabrication process and can lead to much larger errors in the estimation of $\kappa$. Our study has allowed assessing the accuracy of the procedure we used for calculating in-field $\kappa_L$ from an electrical characterization of stabilizer specimens extracted from the reacted tapes. In-field measurements remain important when accurate evaluations of the thermal stability of the conductor are needed, especially in case of high-$RRR$ stabilizers.

\subsection{Transverse Thermal Conductivity} \label{Discussion-k_T}
Analytical studies of the transverse thermal conduction in technical superconductors are more complex with respect to the longitudinal case. Generally, $\kappa_T$ cannot be straightforwardly calculated from the thermal conductivities of raw materials, due to the complexity of the conductor architecture. As a consequence, computational models have to be used to evaluate the overall $\kappa_T$ of the conductor. However, the numerical approach relies on different approximations about both the geometry and the physical parameters of the conductor \cite{Stenvall}. In the light of these considerations, the importance of experimental measurements of $\kappa_T$ in technical superconductors results clear. As far as we know, this is the first experimental study of the transverse thermal conductivity in MgB$_2$ conductors.

The experimental $\kappa_T(T)$ curves have been reported in Figure~\ref{Fig3}. $\kappa_T$ increases monotonically in the explored range of temperatures. In Sample~A and Sample~C, $\kappa_T$ assumes values $\lesssim 10$~WK$^{-1}$m$^{-1}$, whilst in Sample~B we have measured values up to about 40~WK$^{-1}$m$^{-1}$. A comparison between transverse and longitudinal data shows that $\kappa_T$ is much smaller than $\kappa_L$ in a same sample. This is a direct consequence of the fact that $\kappa_L$ is mainly determined by the stabilizer's contribution whilst the transverse thermal conduction is also influenced by the other materials of the composite not optimized for the thermal transport. In order to make this concept clearer, let us consider for simplicity a superconducting tape composed of superimposed layers, each one associated with a material of the composite. The transverse thermal transport can be schematized as in the case of electrical resistances connected in series \cite{Bonura3}. It follows that the overall transverse thermal resistance, $R_T$, can be considered as the sum of single-layer resistances:
\begin{equation}\label{kT}
    R_{T}= \sum \frac{1}{\kappa_i}\frac{t_i}{A} \, ,
\end{equation}
where $\kappa_i$ is the thermal conductivity of the $i^{th}$ layer of thickness $t_i$ and surface area $A$. From Equation~\ref{kT}, one can easily deduce that the transverse thermal resistance is dominated by low-$\kappa$ materials in case of layers of comparable thickness.

The thermal conduction in metals is affected by the magnetic field. This is not the case for thermal insulating materials since the electronic contribution to the heat transfer is negligible. Our study of the in-field thermal conduction performed on Sample~A and Sample~B has revealed that $\kappa_T$ is not affected by the field within our experimental accuracy, in contrast with what observed for $\kappa_L$. Since in the previous section we have demonstrated that $\kappa$ of Cu is considerably reduced after the application of a magnetic field, we deduce that the Cu contribution to the overall transverse thermal resistance is negligible in the investigated samples.

We would like to emphasize that our experimental study can represent an important reference for validating the numerical models commonly used for calculating $\kappa_T$ in technical conductors \cite{Stenvall}. However, this goes beyond the aim of this manuscript, which is mostly focused on the experimental determination of the thermal conductivity in MgB$_2$ tapes.

\subsection{Thermal Stability} \label{Discussion-stability}
The study of thermal conduction in technical superconductors can provide important insights into the thermal stability of superconducting devices.
Due to different kinds of disturbances, the temperature of the conductor may locally increase above $T_{CS}$, the temperature at which a sharing of the operating current, $I_{Op}$, between the superconductor and the stabilizer starts. When the perturbed zone is smaller than the so-called minimum propagation zone, it will eventually shrink, whilst if it is larger, it will grow indefinitely, originating a quench. In adiabatic conditions, the length of the minimum propagation zone can be estimated from: $l_{MPZ}\approx \sqrt{2\kappa_{L}(T_{CS}-T_{Op})/(J_{Cu}^2\rho_{Cu})}$, where $T_{Op}$ is the operating temperature, $J_{Cu}$ and $\rho_{Cu}$ the current density and the electrical resistivity of the stabilizer \cite{Wilson}. The minimum quench energy, i.e. the smallest energy able to trigger a quench, is a key parameter in the evaluation of the stability of superconducting devices. In case of superconducting wires or tapes, $MQE\approx l_{MPZ} S_{Tot} \int_{T_{Op}}^{T_{CS}} \! c(T) \, \mathrm{d}T$, where $S_{Tot}$ and $c$ are the cross section area and the volumetric specific heat of the conductor, respectively \cite{Wilson,Iwasa}. Since we have demonstrated that $\kappa_{L}\approx \kappa_{Cu} s_{Cu}$, using the Wiedemann-Franz law we deduce:

\begin{equation}\label{MQE}
    MQE\approx \frac{\kappa_{L} S_{Tot}^2}{I_{Op}} \sqrt{\frac{2 s_{Cu}(T_{CS}-T_{Op})}{LT_{CS}}}  \int_{T_{Op}}^{T_{CS}} \! c(T) \, \mathrm{d}T  \, .
\end{equation}

$c(T)$ values can be calculated from data reported in the literature for raw materials \cite{Bauer,Kemp,van Weeren}, considering that $c(T)=\sum\limits_{i}v_i c_i$, $v_i$ being the volume fraction occupied by the $i^{th}$ material. Considering the geometry of the conductors, $v_i$ values coincide with $s_i$ data listed in Table~\ref{TabSamples2}. $MQE$ evaluated from Equation~\ref{MQE} at $T_{Op}\approx 20$~K, $B= 1$~T, and $I_{Op}=100$~A is reported in Table~\ref{Tab3}. These parameters define typical operating conditions envisaged in cryogen-free MRI magnets. In the calculation, we have supposed $\kappa_{L}(1$T$)\approx\kappa_{L}(0$T), as this approximation has been shown to be valid over a wide range of temperatures for Cu \cite{Bonura1}. $T_{CS}$ has been extracted from critical-current data supplied by Columbus Superconductors \cite{Tropeano}. The calculated $MQE$s span from $\approx6$~mJ to $\approx34$~mJ, values associated with Sample~A and Sample~C, respectively.

\begin{figure}[t]
\centering \includegraphics[width=8 cm]{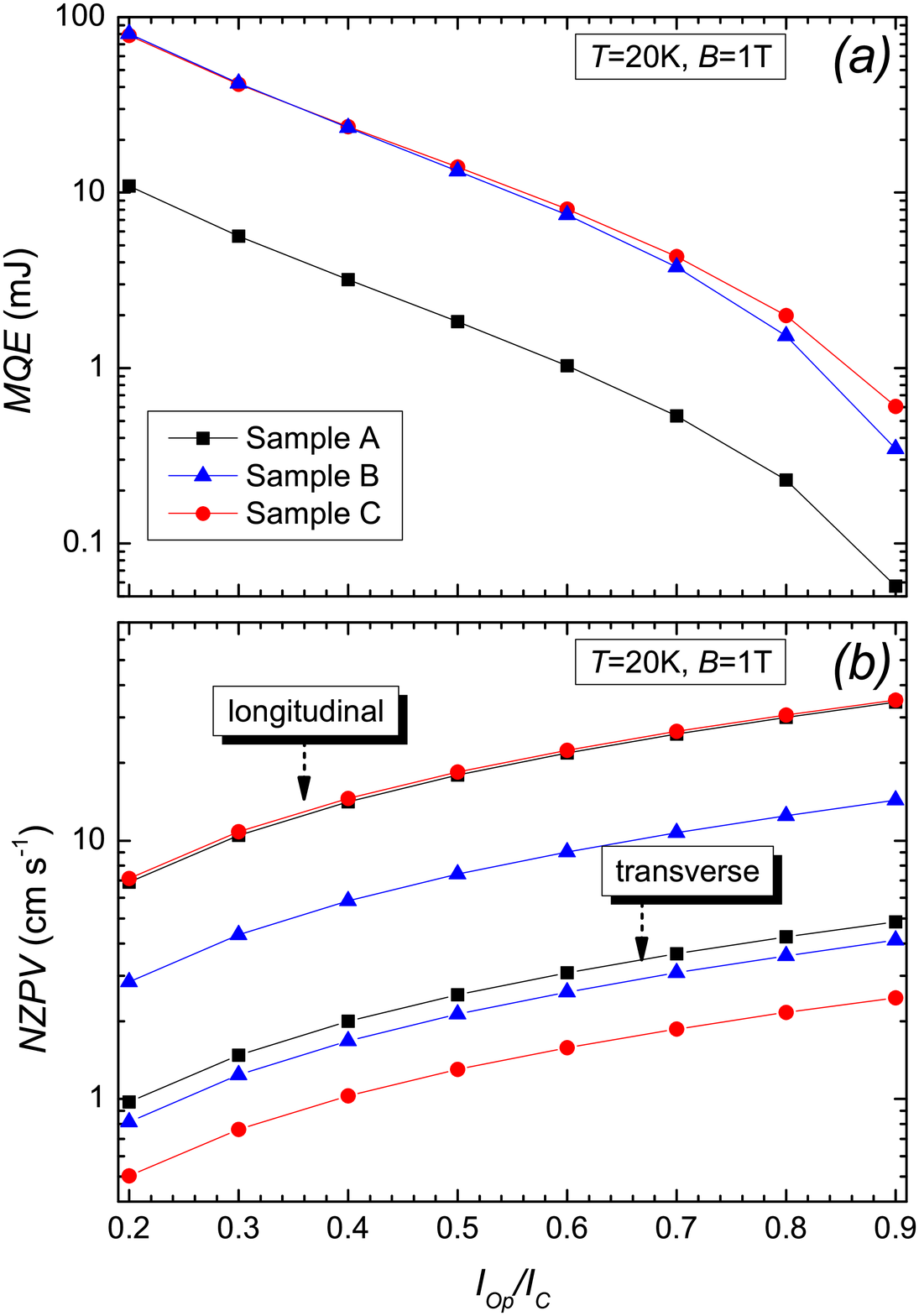} \caption{\label{Fig5}
 Minimum quench energy \textit{(a)} and normal zone propagation velocities \textit{(b)} calculated as described in the text for different values of $I_{Op}/I_C$ at $T=20$~K and $B=1$~T. $I_C\approx345$~A for Sample~A, $I_C\approx240$~A for Sample~B, and $I_C\approx300$~A for Sample~C.}
\end{figure}

The investigated samples differ not only by the thermal properties, but also by their electric behavior. At $T=20$~K, $B=1$~T the critical current is $I_C\approx345$~A, $\approx240$~A and $\approx300$~A in Sample~A, B and C, respectively (see Table~\ref{Tab3}). In order to evaluate the thermal stability of the conductors for the same current margin, in Figure~\ref{Fig5}~(a) we have reported the $MQE$ calculated for different values of the reduced current $I_{Op}/I_C$. The curves associated with Samples~B and C are found to be very close each other. On the contrary, $MQE$ assumes markedly smaller values in Sample~A. The origin of these differences has to be sought in the high $\kappa_L$ values in the case of Sample~C and in the use of stainless steel as matrix material in Sample~B. Indeed, the specific heat of the stainless steel is larger than that of nickel, leading to an increase of the $MQE$ \cite{Bauer,Kemp}.
The $MQE$ values obtained in this work are of the same order of magnitude as the values found experimentally in various MgB$_2$ technical conductors \cite{van Weeren,Martinez,Pelegrin}. Higher $MQE$s of the order of 1~J (at $T\approx20$~K, $B\approx1$~T, $I_{Op}\approx100$~A) have been measured in MgB$_2$ samples with much larger Cu/non-Cu ratios with respect to the tapes investigated in this work \cite{Ye}. However, it is worth noting that $MQE$ of the order of 10~mJ is at least two order of magnitudes higher than typical values of low-$T_c$ superconductors \cite{Iwasa,van Weeren}. Furthermore, large Cu/non-Cu ratios lead to a reduction of the engineering current density.

When designing a superconducting device, one wants to keep the $MQE$ as large as possible. Nevertheless, this is not the only parameter that has to be kept under control. If a quench occurs, the temperature rise has to be limited, otherwise serious damages could occur in the system. This requires a rapid propagation of the perturbation, i.e. large values for the normal zone propagation velocity, $NZPV$. High $NZPV$s are also important to quickly detect the quench and activate the protection system \cite{Park}. In adiabatic conditions, we can use the equation given by Iwasa for estimating the longitudinal component of $NZPV$:
\begin{equation}\label{NZPV}
    NZPV_L\approx \frac{I_{Op}}{S_{Tot}c_{Av}} \sqrt{\frac{\rho\kappa}{T_{S}-T_{Op}}}\approx \frac{I_{Op}}{S_{Tot}c_{Av}} \sqrt{\frac{LT_{S}}{T_{S}-T_{Op}}}  \, ,
\end{equation}
where $c_{Av}=\sqrt{c_sc_n}$ is the geometric average of the volumetric specific heat at $T_{Op}$ ($c_s$) and $T_C$ ($c_n$), and $T_S=(T_{CS}+T_C)/2$ \cite{Iwasa}. We would like to emphasize that, since $NZPV_L$ depends on the product $\rho \kappa$, from the Wiedemann-Franz law one obtains a field-independent expression.

The $NZPV_L$ value calculated by Equation~\ref{NZPV} at $T_{Op}\approx 20$~K and $I_{Op}=100$~A has been listed in Table~\ref{Tab3}. In Figure~\ref{Fig5}~(b) $NZPV_L$ has been plotted for different values of the reduced current. Samples~A and C exhibit comparable propagation velocities, being approximately twice the value calculated for Sample~B. This difference is due to the use of stainless steel instead of nickel in Sample~B, since this leads to a larger $c_{Av}$. $NZPV_L$ values obtained in this work are of the same order of magnitude than those measured experimentally or determined numerically for other MgB$_2$ technical conductors \cite{van Weeren,Martinez,Pelegrin,Ye,Gambardella}.

\begin{figure}[t]
\centering \includegraphics[width=8 cm]{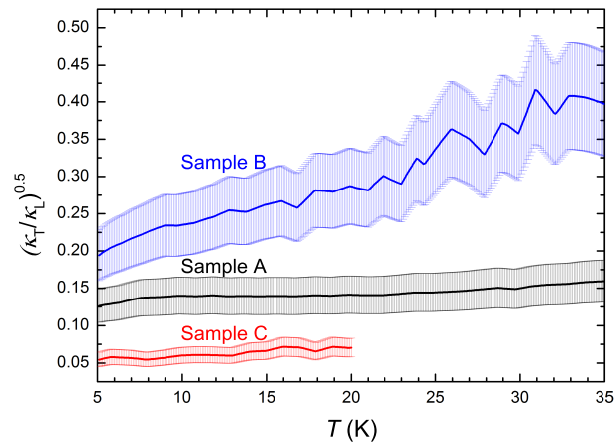} \caption{\label{Fig6}
Temperature dependence of the square root of the ratio $\kappa_T/\kappa_L$.}
\end{figure}

Our simultaneous study of the longitudinal and transverse thermal conductivities has given us the possibility to calculate the anisotropy of the $NZPV$, since $NZPV_T/NZPV_L\approx\sqrt{\kappa_T/\kappa_L}$ \cite{Wilson}. The experimental determination of $NZPV_T$ is much more complex than that of the $NZPV_L$ \cite{Bonura3}. As a consequence, these data represent valuable information to consolidate the numerical models used for predicting the 3D quench propagation in a coil. In Figure~\ref{Fig6} we show the temperature dependence of $\sqrt{\kappa_T / \kappa_L}$ deduced from $\kappa$ data measured at $B=0$. Error bars have been obtained by propagating the experimental data uncertainties.  The largest anisotropy of the $NZPV$ has been found for the Sample C. This origins from the use of high-$RRR$ Cu, since this affects considerably the longitudinal heat conduction, whilst it is not important in the transverse direction. From the experimental values of $\sqrt{\kappa_T/\kappa_L}$ and the $NZPV_L$ data calculated by Equation~\ref{NZPV}, we have deduced the transverse component of $NZPV$ for different values of the reduced current. These data are also presented in Figure~\ref{Fig5}~(b).

From the study of the thermal properties, one can deduce valuable information about the correlation between conductor layout and thermal stability. The use of Cu with high $RRR$ is very effective in increasing $MQE$, especially in case of applications at relatively low magnetic fields. This originates from the fact that $MQE\propto \kappa$. The purity of the stabilizer is not important in determining the propagation velocity of the perturbation because $NZPV$ is proportional to the product $\kappa\rho$, i.e. to the Lorenz number, rather than to $\kappa$. This is why we have found comparable $NZPV$s in Samples A and C, which have similar constituent materials (see Table~\ref{TabSamples2}) but very different $RRR$ values of the stabilizer (see Table~\ref{Tab3}). With regard to the conductor matrix, the use of stainless steel in place of nickel has a double effect on thermal stability: on one hand, it increases $MQE$ and thus the stability of the conductor; on the other, it lowers $NZPV$. The two effects are due to the fact that the specific heat of stainless steel is larger than that of nickel. From our study, we also deduce that the different procedures adopted to stabilize the conductors lead to different $RRR$ values. In particular, soldering a Cu strip on the reacted tape was found to be very effective in maintaining a high $RRR$, because mechanical deformation or poisoning of the stabilizer during the heat treatment are avoided.

\section{Conclusion}
This paper has presented an experimental study of the thermal conduction properties of commercial MgB$_2$ conductors. The longitudinal and transverse components of the thermal conductivity have been measured and the field-induced effects on the thermal conduction explored. The investigated conductors differ in constituent materials and type of stabilization. This has allowed us to correlate thermal properties and conductor architecture.

Our results show that the longitudinal thermal conductivity at low field can vary over one order of magnitude because of the different conductor layout. In particular, the highest $\kappa_L$ values have been measured in a conductor stabilized by soldering a Cu strip on one face of the tape. Indeed, this protects Cu from contamination and mechanical deformations during the fabrication route. $\kappa_L$ at $B=0$ can be estimated with an uncertainty of $\pm20\%$ from the $RRR$ of Cu and the Cu/non-Cu ratio, if the $RRR$ of the stabilizer is evaluated by measuring the $R(T)$ curve in Cu specimens extracted from the conductors. In presence of magnetic fields, $\kappa_L$ is reduced by an extent that depends on the $RRR$ and on the reciprocal orientation between field and thermal current. The accuracy of the procedure for calculating in-field $\kappa_L$ from the magnetoresistance of Cu depends on the purity of the stabilizer. Furthermore, it has been presented the first experimental study on MgB$_2$ of the thermal conduction in the direction normal to the tape. It has been found that transverse thermal diffusion is dominated by non-conducting layers and is field-independent within our experimental accuracy.

The investigation of the thermal transport properties has allowed us to infer about the thermal stability of the conductors, using the concept of minimum quench energy and normal zone propagation velocity. The stability in case of local disturbances can be improved either by using high-$RRR$ Cu or by increasing the volumetric specific heat of the conductor, choosing appropriate materials for the matrix. On the other hand, high values for the volumetric specific heat have the effect of lowering the normal zone propagation velocity.

\section*{Acknowledgment}
We warmly acknowledge Dr. Matteo Tropeano from Columbus Superconductors for his collaboration in this work. Furthermore, we thank Damien Zurmuehle for technical support.

\end{document}